# The canonical idea of Dirac as an alternative to Bohr's approach: A toy model of QDism vs. QBism.

Artur Szczepański*

**Key words:** Quantum foundations, quantum nonlocality, relativistic behavior, topological defects

**PACS:** 03.65.Ta, 01.70.+w, 03.65.Ud, 03.67.-a, 03.57.Bg, 03.30.+p, 61.30Dk

Classical objects have been excluded as subjects of the observed quantum properties, and the related problem of quantum objects' nature has been suspended since the early days of Quantum Theory. Recent experiments show that the problem could be reasonably revisited. The outlined model indicates new issues, which could result from following and exploring the canonical idea of Dirac, so the tag QDism. Topological defects in solids are considered as an example. The aim is helping to grasp the underlying pre-theoretical new intuitions, which should replace the old ones attached to the background of classical physics.

## 1. Introduction

Foundations of Quantum Physics have ever been involved in metatheoretical queries. Historically, the related problems while significant in the interpretational debate, did not appear directly in the development of both the predictive power of the theory, and of the applied research [1] until the advent of the contemporary Quantum Optics, and in particular of Quantum Communication. The questions, inter alia, linked to the status of theoretical quantum objects (e.g. wave functions), and to the status of quantum nonlocality, acquired an experimental and an applied meaning (See e.g. [2,3,4,5]). So some of the "quantum conundrum" problems weighting previously in the academic debate only, have emerged in applications, and in the new foundational issues. The recent both commercial and military applications of QSDC (Quantum Secure Direct Communication) in China, i.e. useful information transmitted by the quantum channel exclusively, which followed the experimental confirmation of QSDC [6], require a new insight into the standard relativity – quantum nonlocality relation (see [7] for an in-depth account).

Here I discuss a simple toy model of objects, which are "excited states of the vacuum", according to the canonical idea of Dirac [1]). Both the relativistic and nonclassical properties of the considered model objects result from a common single physical basis, and therefore are physically cohesive. However, it is important to emphasize that the model is clearly too simple to take it as an attempt to provide ready to use model-objects corresponding to real physical quantum-objects. The aim is rather to point at a the emerging new pre-theoretical intuitions, which could contribute to circumvent the limitations of the standard approach resulting from the lack of a proper "subject of quantum properties". "Peaceful coexistence" [2]) and QBism [9] (as a refinement of the standard approach), could be evoked in the context of the mentioned limitations. "QDism" shall be the tag used in our approach.

---

* Present address: artallszczep@gmail.com (Ret. Form the Inst. of Fund. Research, Pol.Acad. of Sci. Warsaw)

[1]) I do not evoke the existing numerous attempts at developing Dirac's idea, since my aim is to shed some light on the yet unexploited issues.

Note that the often silently accepted "pre-theoretical intuitions" could play a key role in the choice of the (foundation) theory's axioms. Therefore we aim at articulating them expressis verbis.

While the focus is on the foundations, the model is of interest to quantum communication applications: useful classical-channel-free quantum communication is feasible within the model's framework.

The paper is organized as follows: Section 2. is a concise presentation of the model.
In **Appendix 1**. some meta-comments placing the model in a more general context are discussed. **Appendix 2.** contains proposals of feasible classical-channel-free "matter of principle" communication experiments.

## 2. The model

Consider topological point defects in a locally regular crystal matrix: vacancies (V) and interstitial atoms (A). V and A are created pairwise: an atom is hit out from a lattice node and placed in an interstitial position. The emptied node is the center of the vacancy. The dual process, that is the annihilation of the A-V pair consists in filling the vacancy by a lattice atom, which restores the regular lattice structure. The pair-creation energy stored in the perturbed structure is subsequently released as lattice vibrations. The term "lattice atom" denotes in the present context an "elementary" lattice-object, the structure of which is irrelevant to the defect's properties.

Both V and A are ***structural objects***, (SO), understood as definite configurations of the displaced structure of the matrix. The defining state of A and V is the one of nodes maximal displacements symmetry, and shall be referred to as the static equilibrium state, (SES), of the defect. Note that the identity of a SO does not depend on the identity of matrix's atoms involved in the displacement field of the actual defect.

The considered SO is different from a classical physical object : It should be taken neither for an object persisting in time (e.g. a corpuscle), nor for a process (e.g. a wave). For more on the accepted terminology see **Appendix 1.**

In the macro-description the lattice is modeled by a continuous medium. Here we are interested in a medium, which can be treated as homogeneous, isotropic, and incompressible. The incompressibility condition excludes the occurrence of longitudinal (displacement) waves in the medium, which opens the way for a **frictionless free propagation** process of both A and V [10].

---

[2]) The term has been coined by Abner Shimony [8] to describe the relative status of relativity and quantum nonlocality.

Now, let us ask what could be the physics made by a "defect observer", (DO), that is an observer belonging to the "world of defects" [3]). "Making physics", is understood here, grosso modo, in the standard way: formulating theories accounting for the results of observations. Hence the question: what observations are accessible to a DO, that is what kind of interactions feels a DO. The obvious answer is: interactions modifying the physical state of the DO.
A single DO placed in an otherwise unperturbed matrix is a free object since there is nothing which could act on his structure. He would than see an empty space. Therefore the ***unperturbed matrix is "the vacuum" to a DO***: he has no direct experimental (observational) access to the underlying "world of the matrix's nodes-atoms".

Let us refer to the DO's physics based on the above indicated approach to as the "eigenphysics of defects". Since to a DO "physical objects" are "structural perturbations" of the underlying matrix, the notion of "physical matter" in the considered here model acquires a two-layer structure: the layer of "true matter" consisting of matrix atoms, and the layer of defects (see, **Appendix 1.**). There may be matrixes allowing for eigenphysics in which a DO is unable to observe his free motion relative to the vacuum i.e. to the matrix (See, **Appendix 1.** and [10]). He can see free motions relative to other disturbed states of the matrix only, and in particular, to other DOs. Suppose this to lead the DO to assume the kinematical equivalence of reference systems linked to free-moving defects, and consequently to formulate a relativity principle in his eigenphysics.

The next step a DO could do is to conclude from the isotropy of the observed space, and from his relativity principle, that the only admissible kinematical groups in his eigenphysics are the Galilei group and the Lorentz group [11]. But he would exclude the Galilei group since the relative velocity of defects, and consequently of reference systems in his eigenphysics is bounded by the transverse (equivoluminal) wave velocity, $\boldsymbol{v_t}$ .
Thus a DO would take the Lorentz group with the invariant limiting velocity $\boldsymbol{v_t}$ for the kinematical group in his eigenphysics.

We turn back to the micro-level. The defect sees the perturbations of structure only, since he has no direct observational access to the mechanics of the nodes-level. This entails nonclassical effects in the eigenphysics of defects:

**1**. The time-evolution of a defect's state is the time-evolution of the node displacements forming the defects' structure. The process is governed by the (invisible to the defects) mechanics of the lattice nodes, or in other words, by causes which do not belong to the world of defects. So to a DO the observable time evolution contains an essentially invisible and unknowable contribution, which could lead to the appearance of an irreducible "true randomness" [12] in the eigenphysics of defects.
**2**. Consider defects' states consisting of two parts which are spacelike separated in the eigenphysics. Suppose that a modification of the substate in one of the parts triggers a modification of the matrix's state resulting in the change of the second part's substate. Both the modifications of partial states are observable in the DO's eigenphysics. However their physical link, that is the interaction transfer in the underlying matrix, is unobservable to the

[3]) The terms "defect observer" and "world of defects" should be taken for self-clarifying heuristic hints.

DO, since it is not transmitted via matrix's displacement fields, or generally via fields felt by defects. So such an unobservable action is nonlocal to a DO because of lack of intermediary, and furthermore it may appear as superluminal, since $\boldsymbol{v_t}$. is not the limiting velocity at the matrix's level.

Both ad. **1** and ad. **2** are examples of egzocosmic (relative to eigenphysics) actions, the effects of which are observable to a DO. Note that egzocosmic actions are not linked to "hidden parameters" in the sense of the historical quantum foundations debate. The latter have been taken for endocosmic relative to the considered world of objects. The occurrence of egzocosmic physical effects is excluded in a single-layer material world, since there is no place for physical reality outside the material world. This would imply the nonexistence of egzocosmicity in a world of classical objects.

**3.** Note furthermore that the properties of a V (A) in the SES state do not differ from the properties of another V (A) in his SES state. This could be understood as a "structural identity indistiguishability" of defects at the matrix level, but to a DO, i.e. in the DO's eigenphysics, it could appear as "true indistiguishability".

## Appendix 1.

The relation {Physical Theory ↔ Real Physical Objects} in classical physics ({CTh ↔CO}), and in particular in Newtonian mechanics seems to be natural and well-understood at both the foundational and every-day research level. Here I comment on the structure of real classical physical objects, and the structure of the corresponding model objects, to state expressis verbis some (often) silent assumptions needed to make the {CTh↔CO} so smoothly working. I than try to see how analogous assumptions work in our toy-model.

Restating the latter proposal in the historical context of the XX century thirties would mean:

1. Taking the Dirac sea for a solid medium.
2. Treating matter and antimatter on the same footing. Since the antimatter has been formed out of holes in the sea, and since holes are "structural objects", matter should be taken for structural objects, too.

Let us take {CTh↔CO} for a three-component structure consisting of :

1. The formalism of the theory, (F).
2. The theoretical model objects (MO).
3. The classical physical objects (CO).

In standard theories F is based on "sound mathematics", so F is the clearest part of the relation. The MO are axiomatically endowed with a set of basic properties. At a suitable level of generality the results of F applied to the MO-axioms establish new properties of MO. The so deduced properties are interpreted as the predictions of the theory. The latter are tested via (experimental) observations, and the consistency of the observed CO's properties with the predicted MO's properties delimits the useful working range of {CTh↔CO}, that is (provisionally) establish the set physical objects to which properties the classical predictions apply.

The just sketched apparently obvious scheme works so naturally since the MO are additionally endowed with basic features which do not belong to the substructure of properties. To see it clearly we have to turn to the structure of a classical object. In the latter we may distinguish, according to Ingarden [13], two other substructures: the one referred to as the categorial form, and the second denoted as the mode of being.

The classical physical objects of interest have the categorial form of either an object persisting in time (e.g. corpuscles, bodies, media, etc.), or of a process (e.g. motions of objects persisting in time). So we have to do with either properties of objects persisting in time, or properties of processes. The consistency of a MO's properties with the properties of the corresponding classical object can be achieved only, if the content of the MO is endowed with the categorial form of the corresponding classical object. For example a MO corresponding to a planet in Newton's theory is a mass point. Both have the categorial form of an object persisting in time. Both the orbital motion of a planet and the model-motion of the corresponding theoretical mass point have the categorial form of a process.

Classical physical objects are supposed to be existentially independent of our conscious actions, and in particular of our cognitive actions aimed at those objects. This is the main if not the only existential requirement needed to interpret experimental observations as observations of the real world.

Both the "categorial form" and the "existential autonomy" (for the english terminology see e.g. "Roman Ingarden" in the Stanford Encyclopedia of Philosophy) have been so natural and obvious to classical physicists that they have been assumed silently. Note, however, that the MO themselves are theoretical constructs created by our conscious acts. On the other hand the content of their endowment i.e. the properties, the categorial form, and the existential mode once attributed persists in the sense that any modification requires the construction of a new MO. Otherwise the correspondence of the MO and the real objects would not be consistent.

Proceeding accordingly to the just outlined scheme has been regarded in the pre-quantum period as the major contribution to the progress of scientific knowledge concerning the physical reality. In practice the progress consists in gathering previously unknown properties of already known objects, or in revealing the presence of new objects (persisting in time, or processes) and properties of the latter. The role of the remaining parts of the objects' structure (the silently assumed categorial form and mode of being) have been usually unnoticed.

Let us turn to our toy-model. To "physicists of the matrix world" the defects' properties are just a subset of properties belonging to the matrix system. Therefore their properties belong to well-defined subjects, i.e. to objects with easy to grasp and well-defined categorial form, and "existing autonomously". So model objects can be constructed, and the "matrix physicist" may proceed within the three-component scheme: {Theory – Model Objects – Matrix}.

To a physicist living in the “defect world” (say, a DO of the main the text) observable properties would be consistent, if he admit that his body is a structural object. Note that the classical physicist is inclined to take his body for a “true matter” object. So the above stated basic assumption made by the “defect physicist” could be to him a strong and bold assumption. Note that available metatheory does not provide us with tools to infer the categorial form of an object from its properties. This circumstance contribute to the boldness of the DO’s assumption.

The structure-object-assumption, (SOA), would open new communication prospects to the “defect physicist”: In his “defect world” there would occur exceptions to the non-signaling limitation (see, e.g. [14, 15] and more recent papers by Peres). The mentioned exceptions would be of major interest to applications.

The general importance of SOA to the foundations and metatheory would consist in providing the answer to the question about the nature (the categorial form) of the objects existing and forming the material world in which the “defect physicist” is living.

Let me conclude with the following remark: It’s time for the quantum physicist to look for a reasonable categorial form of quantum objects, that is revisit the fundamental question postponed in the early days of Quantum Theory.

## Appendix 2.

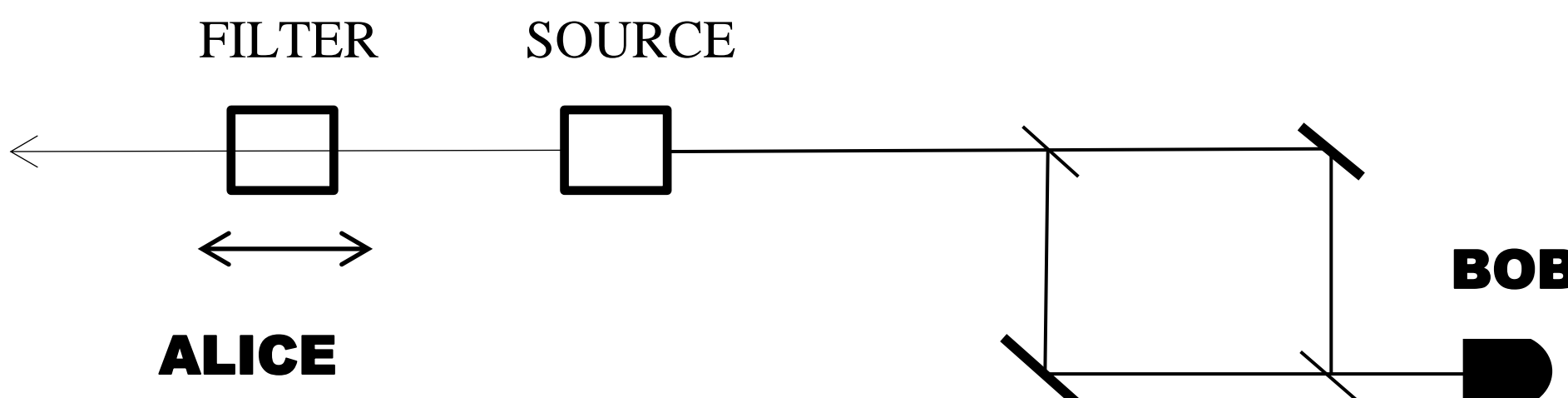

**Fig. 1. Scheme of a steering experiment with energy-entangled biphotons**. One photon (the A-photon) is emitted by the SOURCE into the Alice’s spatial mode, its twin, the B-photon goes into Bob’s spatial mode. Alice controls the length of the optical path between the SOURCE and the FILTER. The latter cuts off a part of the A-photon’s frequency spectrum (projective action). Bob’s detector sees the interference pattern of his Mach-Zehnder interferometer (energy-decoherence-free measurement). The modification of the A-spectrum results in an appropriate B-spectrum modification (steering effect), which can be observed by Bob in the interference pattern. Bright entangled photons sources and fiber interferometry makes the experiment feasible. Note that the experiment is a simplified version of the experiment by Kwiat and Chiao [16], the pioneering importance of which has acquired s full significance in view of [6].

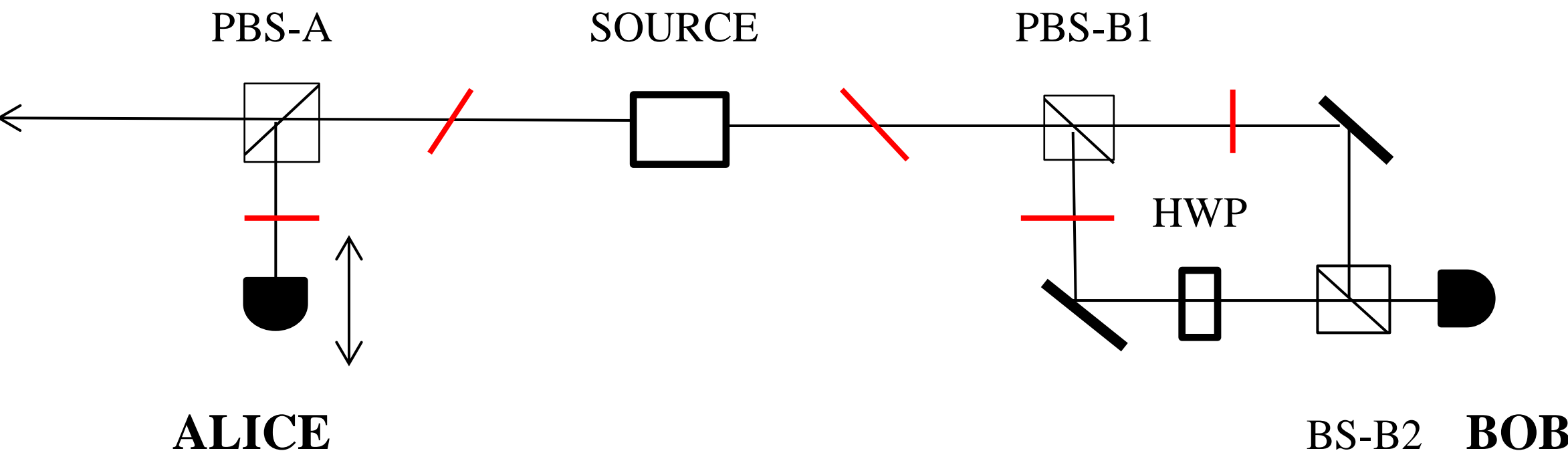

**Fig. 2. Scheme of a steering experiment with polarization-entangled photons.** The linearly polarized A-photon is split by Alice's, and the B-photon is split by Bob's polarizing beam splitter (PBS) into the vertical and horizontal components, respectively. The half wave plate HWP make both the split parts of the B-photon coplanar. BS-B2 is a non-polarizing beam splitter. The interference pattern observed by Bob is a decoherence-free measurement. But if Alice's projective action precedes Bob's measurement his result contributes to a flat pattern. So Bob does not need any classical channel to observe the steering effect.

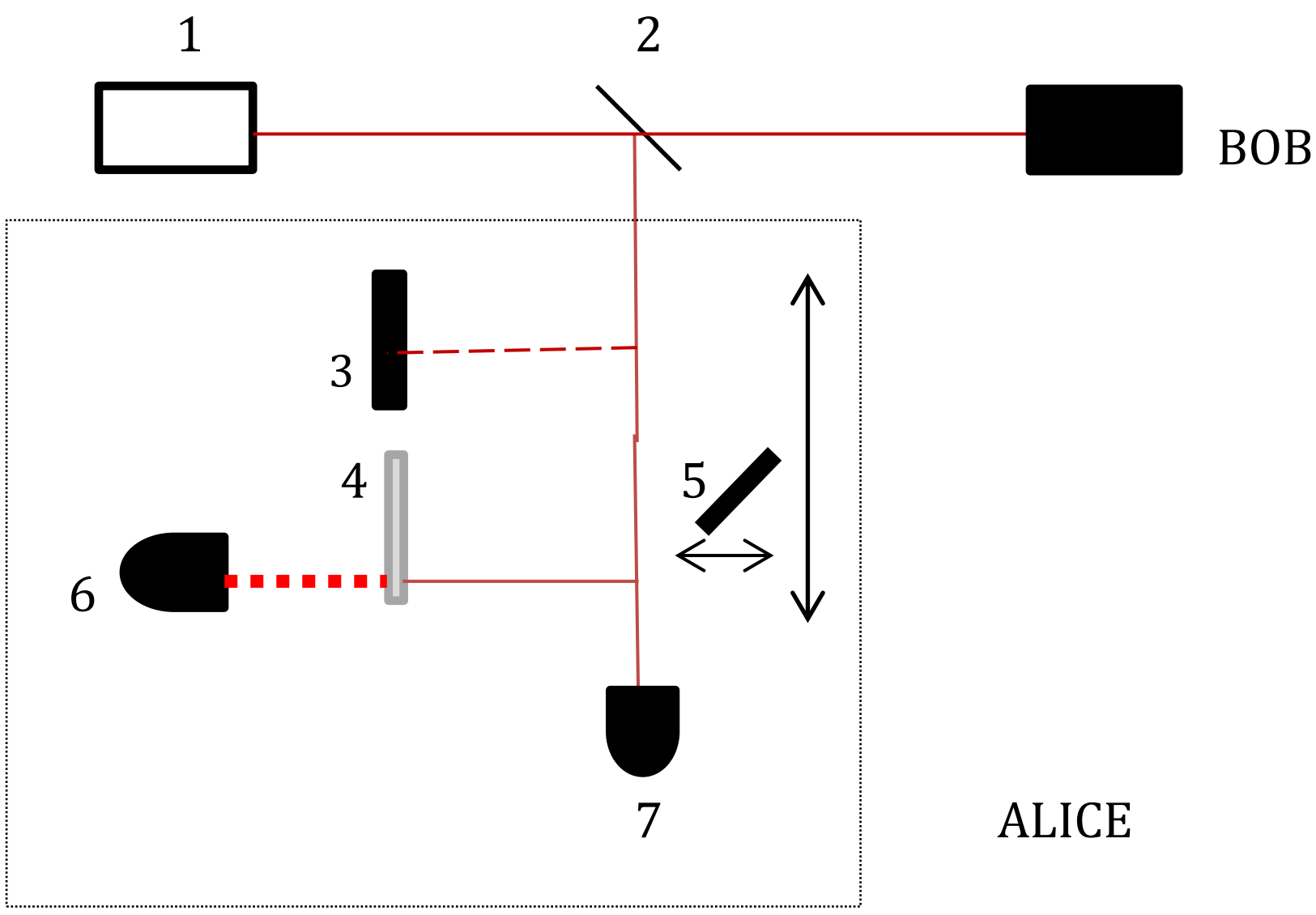

**fig. 3. scheme of a quantum direct communication experiment based on photon statistics modification.** the input laser beam from (1) is split by the beam splitter (2). The transmitted part is directed to bob, and the reflected part is directed to the remaining elements operated by alice: (3) is a black screen, (4) a simple beam decohering device turning the laser beam into para-thermal light, (5) a reflecting mirror, (6) and (7) Alice's measurement devices.

In the first step BOB and Alice (by means of (7)) measure the photon statistics in their beams (e.g. the second statistical moment). The aim is to find the optimal intensity of the input beam relative to the expected results of the final step.

The second step consists in repeating the first step measurement by BOB with mirror (5) directing the beam toward the black screen (3). BOB sets the length of the optical path between him and (2) to exceed the length between (2) and (3):

|BOB – (2)| > |(2) – (3)|. (**L**1)

In the third step ALICE directs the beam toward (4) and measures the statistics by (6). The distances are:

|BOB – (2)| > |(2) – (4)|. (**L2**)

The final step consists in repeating the measurements (**L1**) and (**L2**) width reversed distances:

|BOB – (2)| < |(2) – (3)|, (**S1**)

|BOB – (2)| < |(2) – (4)|. (**S2**)

The direct quantum communication would be achieved if the results (**L1**) and (**S1**) can be interpreted as indistinguishable, while (**L2**) and (**S2**) as different, and (**L1**), (**L2**) as different too.

The reason for expecting such a positive result: The statistics of photons which interacted with (4) and are absorbed by ALICE is that of a para-thermal light. So the statistics of the correspondent breaks in the set of BOB's counts is no longer that of the previous laser light. But the statistics of breaks in a set of counts is just a dual aspect relative to the statistics of the counts themselves.

The variable distance of Alice's projecting action from the SOURCE allows for determining the "propagation velocity of the steering effect", ($V_S$) in the lab's frame, in both the proposed experiments:

$V_S$ = (threshold optical AB-paths difference) / (threshold AB-photons-time of flight difference).

Here, "threshold" means the Alice's action (FILTER or detector) optical distance from the SOURCE at which Bob observes the transition of the interference pattern from unmodified to modified form.

**Acknowledgments**

I am grateful to the participants of the ZTOC seminar at the Institute of Fundamental Technological Research (Warsaw), to the participants of the seminar “Open Systems” at the Chair of Mathematical Methods of Physics of the Warsaw University, and to Michele Caponigro for comments and suggestions. Jakub Rembieliński, and the paticipants of the seminary at the Chair of Theoretical Physics, University of Łódź are acknowledged for comments on the proposed experiments. I am thankful to Małgorzata Felicka for inspiring comments on the two-layer concept of physical world, and especially to Jan Mostowski for interest and encouragement.